\documentclass[aps,prd,reprint,nofootinbib,showkeys,showpacs,shttps://www.overleaf.com/project/5e986462b4e77200012577ffuperscriptaddress,longbibliography,floatfix, twocolumn]{revtex4-1}

\usepackage{ragged2e}

\usepackage{graphicx}
\usepackage{color, colortbl}
\definecolor{Gray}{gray}{0.9}
\definecolor{LightCyan}{rgb}{0.88,1,1}
\usepackage{caption}
\usepackage{subcaption}
\usepackage{amsmath}
\usepackage{amssymb}

\usepackage[american]{babel}
\usepackage{booktabs}
\usepackage{dcolumn}        
\usepackage[T1]{fontenc}    
\usepackage{graphicx}       
\usepackage[dvipsnames]{xcolor}
\usepackage{hyperref}
\hypersetup{
    colorlinks = true,
    linkcolor= blue,
    citecolor = blue,
    urlcolor = blue,
}
\usepackage[utf8]{inputenc}        
\usepackage{multirow}
\usepackage{txfonts}               
\usepackage{url}
\usepackage{float}

\usepackage{lipsum}
\newcolumntype{d}[1]{D{.}{.}{#1}}
\newcolumntype{v}[1]{D{,}{,\ }{#1}}

\graphicspath{{Figuras/}}

\providecommand{\keywords}[1]
{
  \small	
   {\textit{Keywords---}} #1
}
\hypersetup{
citecolor = blue,
linkcolor = RoyalPurple,
urlcolor = blue
}

\begin{document}

\title{Non-Parametric Analysis for the Dark Matter Density Evolution}

\author{Z. C. Santana$^{1}$} \email{zilmarjunior@hotmail.com.br}
\author{R. F. L. Holanda$^{1}$} \email{holandarfl@fisica.ufrn.br}
\author{R. Silva$^{1,2}$} \email{raimundosilva@fisica.ufrn.br}

\affiliation{$^{1}$Departamento de F\'{\i}sica, Federal University of Rio Grande do Norte, Natal-RN, 59072-970, Brazil}
\affiliation{$^{2}$Departamento de F\'{\i}sica, Universidade do Estado do Rio Grande do Norte,  Mossor\'o-RN, 59610-210, Brazil}








\date{\today}

\begin{abstract}
In this paper, we investigate a potential departure in the standard dark matter density evolution law, $\rho_{dm} = \rho_{dm,0}(1+z)^3$. The method involves considering a deformed evolution model, denoted as $\rho_{dm} = \rho_{dm,0}(1+z)^3f(z)$, and searching the presence of any deviation ($f(z)\neq 1$). As one may see, $f(z)$ is a general function that parametrizes a possible digression from the standard law. We use data of baryon acoustic oscillations, type I Supernovae luminosity distances, and galaxy cluster gas mass fraction observations to reconstruct $f(z)$ through an approach that is not dependent on the cosmological model or the so-called Gaussian process regression. Unlike previous works, it enables us to investigate a  possible deviation without using a specific function to describe it. We have obtained $f(z)=1$, the standard model scenario, within $2\sigma$ c.l. in all the considered cases.
\end{abstract}

\maketitle

\section{Introduction}\label{sec:intro}

Since the discovery of accelerated expansion \cite{riess1998observational, perlmutter1999measurements, weinberg2013observational}, the physical origin of the mechanism driving this effect remains unknown \cite{weinberg2013observational}. In this context, the dark energy is the primary component that characterizes the standard model, known as $\mathrm{\Lambda}$CDM, followed by dark matter, which is responsible for the formation and dynamics of large-scale structures in the universe \cite{ozer1986possible, sahni2005dark, caldera2009growth}. The standard model exhibits good predictive power and agrees with observations, such as the cosmic microwave background observations by the Planck satellite \cite{ade2016planck}. However, despite its robustness, there are several observational discrepancies, including the tension in $H_0$ \cite{kamionkowski2023hubble, di2021realm}, the cosmological constant problem \cite{weinberg1989cosmological, padmanabhan2003cosmological, lombriser2023cosmology}, the cosmic coincidence problem \cite{zlatev1999quintessence}, among others (for a review of some tensions see \cite{di2021realm, perivolaropoulos2022challenges}). Therefore, exploring alternative theories to the standard model is imperative to address these observational discrepancies and advance our understanding of the universe's fundamental principles \cite{elcio2022cosmology}.

A range of proposals to address these issues involve considering extensions to the standard model \cite{joyce2015beyond}, allowing, for example, for non-gravitational interactions between the dark sector components \cite{wang2016dark}. In this scenario, the dark matter and dark energy dominating the current evolution of the universe interact with each other. As their densities coevolve, a natural explanation for the coincidence problem is provided (See \cite{amendola2000coupled} and references therein). {The ground of this scenario is based on the cosmological constant problem (see \cite{carroll2001cosmological, weinberg1989cosmological} for reviews). Strikingly, the so-called cosmological constant problem is associated with understanding the mechanism that provides small vacuum energy in the same order of magnitude as the present matter density of the universe \cite{perlmutter1999measurements}. Based on the quantum field theory, some approaches have successfully addressed the dynamical cosmological constant \cite{nelson1982scaling, elizalde1994renormalization, elizalde1995gut, bytsenko1994effective, wagoner1970scalar, linde1974lee, polyakov1982phase, cohen1999effective, shapiro2002scaling, shapiro2000scaling, shapiro2003variable, espana2004testing, weinberg1993vacuum,sola2011cosmologies}.}

From the phenomenological standpoint, many interaction models have been proposed; however, each model is entirely independent. Theoretically, the second law of thermodynamics and the Le Chatelier-Braun principle impose constraints on how these components interact, suggesting that dark energy decays into dark matter \cite{alcaniz2005interpreting}. Moreover, in the opposite scenario, the cosmic coincidence problem is exacerbated \cite{jesus2022can}. Although, \cite{pereira2009can} show that assuming both fluids have different temperatures, if at least one has a non-vanishing chemical potential, the decay of dark matter into dark energy is possible. However, there is also no consensus, with results suggesting both scenarios \cite{wang2016dark}. 
Despite the theoretical issues,  a common approach is to consider phenomenological models of interaction and explore their consequences for cosmological dynamics by comparing them with observational data \cite{wang2016dark,bolotin2015cosmological,amendola2007consequences,tamanini2015phenomenological}. Many models following phenomenological arguments are observationally tested in order to capture the possible interactions among dark components, e.g., in \cite{nunes2022new}, a strong constraint on the interaction is imposed through a joint analysis of the galaxy power spectrum, Baryonic Acoustic Oscillations (BAO), and Cosmic Microwave Background (CMB), this study obtains a constraint on the interaction parameter of $|\epsilon| \lesssim 0.1$ for the considered interaction model; in \cite{yang2019dark}, BAO data indicate a preference for  $\epsilon(z) > 0$, implying a lower value for the matter density today; the tension in $H_0$ is alleviated in \cite{lucca2020shedding} by employing an interaction model in analysis with Planck, BAO, and Pantheon data.   It is worth commenting that very recently, based on a set of $N$-body simulations, the authors from the Ref. \cite{liu2022dark} investigated the formation histories and properties of dark matter haloes in scenarios where such interaction occurs and compared with their $\mathrm{\Lambda}$CDM counterparts. Their results showed that dark matter halo formation can be significantly affected by a possible interaction of the two dark components. Therefore, the constraints from non-linear structures are indispensable, the evolution of the gas fraction being a necessary ingredient in the description of the hierarchical growth of clusters.

Other recent works include another approach by investigating a possible variation in the evolution law of matter density caused by some interaction between the components of the dark sector.
In the phenomenological view, the evolution law of matter density is modified to be $\propto(1+z)^{3 + \epsilon}$ \cite{wang2004can} (if $\epsilon = 0$ the standard relation is recovered). For instance, in \cite{bora2022test},  galaxy cluster data from Chandra X-ray observations and cosmic chronometers were used to put constraints on $\epsilon$, while in \cite{holanda2019estimate} Type Ia supernovae data were considered. In \cite{bora2021probing}, in addition to gas mass fraction data,  strong gravitational lensing data are used (obtained from a combination of SLOAN Lens ACS, BOSS Emission-line Lens Survey, Strong Legacy Survey SL2S along with systems discovered by SLACS) to impose constraints on the same relation. As general results, these works indicated a non-interaction scenario, but not sufficient to discard this possibility entirely. 

In this work,  we use data of baryon acoustic oscillations within the redshift range of $0.11 \leq z \leq 2.4$, type I Supernovae with redshift in the range $0.01 \leq z \leq 1.914$ and galaxy cluster gas mass fraction observations with redshifts in the range of $0.0473 \leq z \leq 1.235$ to propose a new approach to investigate a possible deviation in the dark matter density evolution  standard law. However, unlike \cite{holanda2019estimate, bora2021probing, bora2022test}, no specific function is used to parameterize such possible deviation. In this hypothesis, the evolution law of dark matter density is considered to be $\rho_{dm} = \rho_{dm,0}(1+z)^3f(z)$ and the $f(z)$ quantity is  {reconstructed using Gaussian process regression (see more in the Section \ref{sec5}).}

Our paper is organized as follows: Section II briefly discusses the interacting model. Section III presents a brief theoretical review of the adopted methodology, while Section IV describes the dataset used in this letter. The statistical analysis is described in Section V. In Section VI, we present the results, and in Section VII, we provide a summary of the study's main conclusions.

\section{ {Interacting Model}}\label{sec:Interacting_Model}

Considering a homogeneous, isotropic, and flat cosmological background described by the Friedmann-Lemaître-Robertson-Walker metric (FLRW) and assuming that the cosmic budget is composed of baryons (b), dark matter (dm), radiation (r), and dark energy (de). The interacting model considers the dark matter and dark energy as interacting fluids with the energy-momentum tensor of the dark sector given by
\begin{equation}\label{coupled-tensor}
	T_{\mu\nu} = T_{\mu\nu}^{{dm}} + T_{\mu\nu}^{{de}}.
\end{equation}

The covariant conservation of energy-momentum tensor, $\nabla_{\mu}T^{\mu\nu} = 0$, leads to
\begin{equation}\label{conservation}
\dot{\rho}_{dm} + 3H\rho_{dm} = -\dot{\rho}_{de} - 3H\rho_{de}(1+\omega) = Q,
\end{equation}
where $\rho_{{dm}}$ and $\rho_{{de}}$ represent the energy density of cold dark matter and dark energy, respectively, while $Q$ is the phenomenological interaction term. Note that $Q > 0$ indicates the dark energy decaying into the dark matter while $Q < 0$ implies the opposite. 

The evolution of the dark components can be found by solving the system of Eqs. (\ref{conservation}). Generally, this can be done by assuming a form for $Q$  \cite{wang2004can, alcaniz2005interpreting, von2019cosmological} or by assuming a relation between the energy densities of the components \cite{cid2019bayesian, von2020unphysical}. Since in the standard description, the dark matter density evolves as $\rho_{dm} \propto (1+z)^{3}$, such a model considers a deviation from the standard evolution characterized by the following function \cite{costa2010cosmological, costa2010coupled}
\begin{equation}
\rho_{{dm}}=\rho_{{dm},0}(1+z)^{3} f(z),
\label{Eqrhodmevo}
\end{equation}
where $\rho_{{dm},0}$ is the today dark matter energy density calculated in $z = 0$ and $f(z)=(1+z)^{\epsilon (z)}$ is a function that depends of redshift. Indeed, there are some parametrizations for $\epsilon (z)$ as well as thermodynamical constraints associated with the signal of  the interaction constant parameter, $\epsilon_0$ \cite{da2020thermodynamic}. Specifically,  such thermodynamical constraints under the signal parameter of the interaction arise from fluid description of the dark sector of the universe. The second law of thermodynamics and the positiveness of entropy constrain the interaction constant parameter to be positive, i.e., $\epsilon_0 >0$, with $\epsilon(z)=\epsilon_0 g(z)$ \cite{alcaniz2005interpreting,Gonzalez2018,da2020thermodynamic}. However, considering fluid description and assuming they have different temperatures, the Ref. \cite{pereira2009can} showed the possibility of obtaining a negative value for the interaction parameter through thermodynamics arguments. In this letter, we do not assume a specific parametrization, and we relax these conditions, allowing the data to constrain the possibility of interaction in the dark sector through the Gaussian process regression for $f(z)$.

\section{Methodology}\label{sec3}
	
Galaxy clusters, the largest gravitationally bound structures in the universe, hold significant potential for providing a wealth of cosmological information \cite{allen2011cosmological}. In recent years, their observations have been extensively utilized to conduct various cosmological tests \cite{qiu2023cosmology, chaubal2022improving, mantz2022cosmological, corasaniti2021cosmological, wu2021cosmology, holanda2020low, lesci2022amico}. The gas mass fraction ($f_{\text{gas}}$) provides important insights into the baryonic content and the physical processes occurring within galaxy clusters. It is expected that the mass fraction of galaxy clusters approximately matches the cosmic baryon fraction, $\Omega_b/\Omega_m$, where the subindex $m$ corresponds to total matter \cite{sasaki1996new, allen2008improved, allen2011cosmological, holanda2020low,  mantz2022cosmological}, and then, to be used to constrain cosmological parameters. In our work, we followed the modeling of the $f_{\text{gas}}$  given by \cite{allen2008improved}, where

\begin{equation}
f_{\text{gas}}(z) = K\gamma \Bigg( \frac{\Omega_b}{\Omega_m} \Bigg) \Bigg[ \frac{D_A^{*}(z)}{D_A(z)}\Bigg]^{3/2} - f_{*}.
\label{EqGMF1}
\end{equation}
Here, $D_A$ is the angular diameter distance to the galaxy cluster, $D_A^*$ is the angular diameter distance of the fiducial cosmological model used to infer the gas mass measurement ($H_0 = 70$ km/sec/Mpc, and $\Omega_m = 0.3$ in a flat curvature). $K$ is the calibration constant, which accounts for any inaccuracies in instrument calibration, bias in measured masses due to substructure, bulk motions, and/or non-thermal pressure in the cluster gas \cite{mantz2014cosmology}. $\gamma$ represents the gas depletion factor, a measurement of how much baryonic gas is thermalized within the cluster potential and thereby depleted compared to the cosmic mean \cite{battaglia2013cluster, applegate2016cosmology, holanda2017cosmological}. $\Omega_b$ and $\Omega_m$ are the baryonic and total mass density parameters, respectively. The stellar fraction, $f_*$, represents the proportion of baryonic matter within the cluster that exists in the stellar form, comprising the mass contributed by stars. 


Therefore, taking into account the proposed evolution law, such as $\rho_{dm}=\rho_{dm,0}(1+z)^{3}f(z)$, and that $\Omega_m = \Omega_{dm} + \Omega_b$, the Eq. (\ref{EqGMF1}) can be reformulated in the following manner:

\begin{equation}
f(z)= \Bigg( \frac{\rho_{b,0}}{\rho_{dm,0}} \Bigg) \Bigg[ \Bigg( \frac{K\gamma}{f_{\text{gas}}(z) + f_*} \Bigg)\Bigg( \frac{D_A^*(z)}{D_A(z)} \Bigg)^{3/2} -1 \Bigg].
\label{EqGMF2}
\end{equation}
The regression method via Gaussian processes can reconstruct the $f(z)$ behavior in the Eq. (\ref{EqGMF2}) and, then,  the reconstruction of $f(z)$ becomes independent of a specific function, differently of previous works. As one may see, we have access to almost all terms on the right-hand side, except for the angular diameter distance for each cluster in the sample. {  Therefore, in the first moment, we will obtain the angular diameter distance for each cluster by performing the regression method using BAO data (see the following subsection). The same method can be employed, in a second moment, to obtain $D_A(z)$ for each cluster by using type Ia supernovae and considering the cosmic distance duality relation $(D_A(z)=D_L(z)/(1+z)^2)$. Then, we obtain reconstructions for $f(z)$ through different observables.}

	\section{Dataset}\label{sec4}
        \subsection{Gas mass fraction}

\begin{figure}
    	\includegraphics[width=\columnwidth]{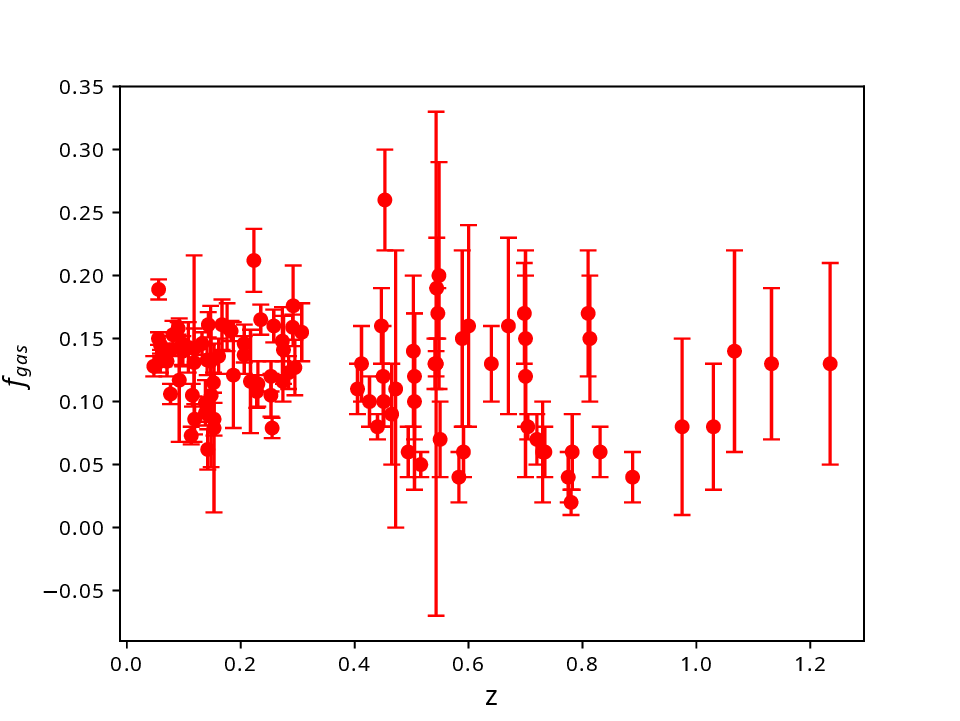}
    		\caption{Measurements of the 103 gas fraction samples used in this study.}
    		\label{datafgas}
    	\end{figure}

            The gas mass fraction observations consist of 103 data on a redshift range of $0.0473 \leq z \leq 1.235$. Estimates of this quantity can be derived by observing and analyzing the temperature and density of the X-ray emissions emanating from the intra-cluster medium within galaxy clusters. Here, our focus lies on the measurements of $f_{\text{gas}}$ within $R_{500}$, which represents the radius of the cluster where the density of the medium exceeds 500 times the critical energy density. This dataset comprises 12 clusters at $z < 0.1$ obtained from X-COP \cite{eckert2019non}; a set of 44 clusters within the range $0.1 \leq z \leq 0.3$ \cite{ettori2010mass}; and observations at high redshifts, consisting of 47 clusters obtained by \cite{ghirardini2017evolution} in the range $0.4 \leq z \leq 1.2$. This dataset was curated by \cite{corasaniti2021cosmological}. It is important to stress that Ref. \cite{corasaniti2021cosmological} did not find a significant redshift evolution to the $\gamma$ factor. {   Generally, the mass of a galaxy cluster obtained through strong gravitational lensing tends to be greater than the mass derived from the hydrostatic equilibrium hypothesis. In other words, the hydrostatic equilibrium assumption used in the sample studies may overestimate the gas mass fraction measurements and, as a result, underestimate the angular distance estimates. However, the galaxy clusters considered in these analyses are limited to the most dynamically relaxed and massive clusters known. This restriction is critical for minimizing systematic errors related to hydrostatic equilibrium and spherical symmetry. Nevertheless, we explore different parameter values $k$ (mass bias) to ensure robust results.} Fig. \ref{datafgas} shows a plot of the  $f_{\text{gas}}$ sample cited.

        In our method,  11 clusters with $z < 0.09$ were utilized to derive constraints for $\rho_{b,0}$ and $\rho_{dm,0}$.  In other words, any constraint obtained with these data is not significantly  sensitive to the nature of dark energy \cite{mantz2014cosmology}. Additionally, we can consider $\frac{D_A^*}{D_A} \approx \frac{h}{h_*}$. Hence, Eq. (\ref{EqGMF1}) can be rewritten as:

        \begin{equation}
            \frac{\Omega_b}{\Omega_m}h^{3/2} = \frac{f_{\text{gas}} + f_*}{K\gamma 0.7^{3/2}}.
        \label{EqVinculos}
        \end{equation}  
       
        By combining this relation with constraints on the reduced Hubble constant and the value obtained from Big Bang nucleosynthesis for the  $100\Omega_{b,0}h^2$ quantity, $100\Omega_{b,0}h^2 = 2.235 \pm 0.033$ \cite{cooke2018one}, it is possible to derive constraints for $\rho_{b,0}$ and $\rho_{dm,0}$ for each combination of $K$ and $\gamma$ used. We chose two options for the reduced Hubble constant value: $h^{\text{SH0ES}} = 0.7324 \pm 0.0174$ (value obtained in the SH0ES project \cite{riess20162}) and $h^{\text{Planck}} = 0.6736 \pm 0.0054$ (from the Planck Collaboration \cite{aghanim2020planck}). Thus, we also cover possible behaviors related to the observational origin of these estimates. Additionally, these values were used to estimate the BAO and supernovae distance data, and we will separate the analyses performed here to maintain consistency in the values of $h$. This type of approximation is possible because of the large dimensions of the considered galaxy clusters and their relaxed nature, and it is assumed that the ratio of baryonic mass to total mass of the cluster is a good representation of the cosmic mean \cite{sasaki1996new}. However, given the different values for $K$ and $\gamma$ used in literature for this galaxy cluster sample \cite{corasaniti2021cosmological}, the baryon to the cold dark matter ratio can differ for each case.

\begin{table}
\centering
\begin{tabular}{|@{}c|c|c|}
\hline
Combination & $\rho_{\text{b},0}(\text{x} 10^{-31}gm/cm^3) $ & $\rho_{\text{dm},0}(\text{x} 10^{-31}gm/cm^3) $ \\
\hline
$K^{\text{CMB}}$ $\gamma^{\text{The300}}$$h^{\text{Planck}}$ & $4.93 \pm 0.732$  &  $13.61 \pm 2.3$ \\
\hline
$K^{\text{CLASH}}$ $\gamma^{\text{The300}}$$h^{\text{Planck}}$ & $4.93 \pm 0.732$  &  $17.42 \pm 3.176$ \\
\hline
$K^{\text{CLASH}}$ $\gamma^{\text{FABLE}}$$h^{\text{Planck}}$ & $4.93 \pm 0.732$  &  $17.25 \pm 3.14$ \\
\hline
$K^{\text{CCCP}}$ $\gamma^{\text{The300}}$$h^{\text{Planck}}$ & $4.93 \pm 0.732$  &  $18.93 \pm 3.072$ \\
\hline
$K^{\text{CMB}}$ $\gamma^{\text{The300}}$$h^{\text{SH0ES}}$ & $4.17 \pm 0.646$  & $13.59 \pm 2.366$ \\
\hline
$K^{\text{CLASH}}$ $\gamma^{\text{The300}}$$h^{\text{SH0ES}}$ & $4.17 \pm 0.646$  &  $17.3 \pm 3.24$  \\
\hline
$K^{\text{CLASH}}$ $\gamma^{\text{FABLE}}$$h^{\text{SH0ES}}$ & $4.17 \pm 0.646$  &  $17.15 \pm 3.204$ \\
\hline
$K^{\text{CCCP}}$ $\gamma^{\text{The300}}$$h^{\text{SH0ES}}$ & $4.17 \pm 0.646$  &  $18.77 \pm 3.182$ \\
\hline
\end{tabular}
\caption{Estimated values at $\rho_{\text{b},0}$ and $\rho_{\text{dm},0}$ obtained from the Eq.(6) for each combination of $K$, $\gamma$ and $h$.  }
\label{Tabvincrhoc0}
\end{table}
        
    As commented earlier, dark matter halo formation  can be significantly affected by
a possible interaction of the two dark components \cite{liu2022dark}. Moreover, the evolution of the gas fraction is clearly a necessary ingredient in the description of the hierarchical growth of clusters.
        \subsection{Baryon acoustic oscillations}

         We utilized a dataset of baryon acoustic oscillations, comprising 18 samples within a redshift range of $0.11 \leq z \leq 2.4$, compiled by \cite{staicova2022constraining} to derive the angular diameter distances of the galaxy clusters. To maintain consistency, we employed two different values for the sound horizon distance, $r_d = 136.1 \pm 2.7$ Mpc, obtained through a late-time estimate using H0LiCOW+SN+BAO+SH0ES \cite{arendse2020cosmic}, and $r_d = 147.09 \pm 0.26$ Mpc, based on Planck results \cite{aghanim2020planck}. Consequently, we obtained two datasets of angular diameter distances. We reconstructed functions for $D_A(z)$ from these datasets using Gaussian process regression, allowing us to estimate this quantity for each galaxy cluster  {and apply it in the Eq. (\ref{EqGMF2}}). Fig. \ref{FigDistancesRecons} shows the angular diameter distance reconstruction function considering both calibrations. It can be observed that the reconstruction adequately covers the data points and exhibits the expected behavior for the angular distance. As shown in this figure, the entire gas mass fraction sample falls within the redshift range of the BAO data, allowing robust estimates of $D_A(z)$.
	\subsection{Type Ia Supernovae}

  Finally, we consider the 1048 data from the Pantheon sample \cite{scolnic2018complete}. It is one of Supernovae's most significant combined samples, with redshift in the range $0.01 \leq z \leq 1.914$. Here, we need to assume a value for the supernova absolute magnitude ($M_B$)
 to turn the distance modulus into measurements of luminosity distance. We consider two values: $M_B=-19.244 \pm 0.037$ (SH0ES team) \cite{camarena2021use} in order to obtain results independent of the cosmological model and $M_B=-19.43 \pm 0.02$ obtained using the Hubble constant estimate by the Planck Collaborations in the context of a $\Lambda$CDM model. The luminosity distances from the reconstruction function for each case can be seen in Fig. \ref{FigDistancesRecons}. Again, the reconstructions adequately covers the galaxy cluster data points and exhibits the expected behavior for the luminosity distance. As commented earlier, one may obtain $D_A(z)$ by using type Ia Supernovae by considering the cosmic distance duality relation $(D_A(z)=D_L(z)/(1+z)^2)$.  {So, using the reconstructions of the $D_L(z)$ functions and the cosmic distance duality relation, we have another method to estimate $D_A(z)$ at the redshift of galaxy clusters and apply it in Eq. (\ref{EqGMF2}).}
	
	\section{Priors and analysis}\label{sec5}
	
	We employed the Markov Chain Monte Carlo (MCMC) method, specifically the Affine-Invariant Ensemble Sampler from the emcee library \cite{foreman2013emcee}, to impose constraints on the parameters of Eq. (\ref{EqVinculos}) and obtain the values for $\rho_{b,0}$ and $\rho_{dm,0}$. The quality of the chains was analyzed using the Gelman-Rubin test \cite{gelman1992inference}, with a threshold value of $ R < 1.1$, and they exhibited rapid convergence. The constraints obtained in this analysis are presented in the Table \ref{Tabvincrhoc0}.

 {In this work, we use Gaussian process regression (GPR) to reconstruct the functions $D_A(z)$ or $D_L(z)$ (from the BAO and SNe Ia datasets, respectively). Therefore, once we have all the values on the right-hand side of Equation
\begin{equation}
f(z)= \Bigg( \frac{\rho_{b,0}}{\rho_{dm,0}} \Bigg) \Bigg[ \Bigg( \frac{K\gamma}{f_{\text{gas}}(z) + f_*} \Bigg)\Bigg( \frac{D_A^*(z)}{D_A(z)} \Bigg)^{3/2} -1 \Bigg],
\label{EqGMF2_2}
\end{equation}
we can again apply Gaussian Process Regression GPR to reconstruct the function $f(z)$, which represents a deviation from the standard dark matter density evolution law for $f(z) \ne 1$. GPR is a machine learning technique used to reconstruct functions in a non-parametric way. Instead of assuming a specific functional form to describe the data, GPR models the function as a probability distribution over all possible functions that could have generated the observed data \cite{williams2006gaussian}. The kernel, or covariance matrix, is crucial in defining how data points are correlated. It allows GPR to infer the underlying behavior of the data and predict new values, constructing a function that flexibly and adaptively represents the data while capturing both the central trend and associated uncertainties. The chosen kernel for both regressions was the Matérn kernel, a reasonably general covariance function multiplied by the constant kernel for scaling. The Matérn kernel is given by
\begin{equation}
    k(x_i, x_j) = \sigma^2\frac{1}{\Gamma(\nu)2^{\nu-1}}\Bigg( \frac{\sqrt{2\nu}}{l}d(x_i,x_j) \Bigg)^{\nu}K_{\nu}\Bigg( \frac{\sqrt{2\nu}}{l}d(x_i,x_j) \Bigg),
    \label{EqKernel}
\end{equation}
where $d(\cdot,\cdot)$ is the euclidian distance between the data, $K_{\nu}(\cdot)$ is a modified Bessel function, and $\Gamma(\cdot)$ is the gamma function. $\sigma$ and $l$ are the variances and the length scale, which control how quickly the correlation between the data changes with the distance, respectively. These are hyperparameters that can be adjusted according to the data.} For the reconstruction of $f(z)$, the length scale was chosen to be $l = 3/2$ to avoid underfitting or overfitting due to the scattered nature of the data and its significant uncertainties.  {The parameter $\nu$ controls the differentiability and smoothness of the reconstructed function. We used in this work $\nu = 3/2$.}

 {In a Gaussian Process, the mean function and the covariance function specify the method entirely. In our work, we used a zero mean function, which assumes that, before observing the data, the function's average value is zero across all inputs. During the regression process, the observed data points influence the mean function, effectively adjusting it according to the patterns observed in the data. Ultimately, this adjusted mean function, combined with the covariance function, allows us to estimate values of the reconstructed function at new points beyond the training set. For recent reviews, see \cite{wang2023intuitive, schulz2018tutorial}.}

An important fact to consider during the reconstruction of $f(z)$ is that for $z=0$, we should have $f(z) = 1$, as can be seen from the Eq. \ref{Eqrhodmevo}. Therefore, this point should be considered as a prior in order to achieve mathematical and physical consistency.

We considered several combinations for $K$ and $\gamma$ obtained from the literature. From the CLASH sample \cite{sereno2015comparing}, we adopted the prior $K^{\text{CLASH}} = 0.78 \pm 0.09$. From the Canadian Cluster Comparison Project (CCCP) \cite{herbonnet2020cccp, hoekstra2015canadian}, we used $K^{\text{CCCP}} = 0.84 \pm 0.04$. A joint analysis of the Planck primary CMB, Planck-SZ number counts, Planck-thermal SZ power spectrum, and BAO provided the value $K^{\text{CMB}} = 0.65 \pm 0.04$ as reported in \cite{salvati2018constraints}. For the depletion factor, we utilized $\gamma^{\text{The300}} = 0.938 \pm 0.041$ inferred from samples of The Three Hundred project \cite{eckert2019non, cui2018three}. Another prior was obtained from the FABLE simulations \cite{henden2020baryon}, where we adopted a constant value of $\gamma^{\text{FABLE}} = 0.931 \pm 0.04$. Finally, the value used for the stellar fraction was $f_* = 0.015 \pm 0.005$ \cite{eckert2019non}.

	\section{Results and discussion}\label{sec6}
 
\begin{figure*}
    \centering
    \begin{subfigure}[b]{0.45\textwidth}
        \centering
        \includegraphics[width=\textwidth]{ 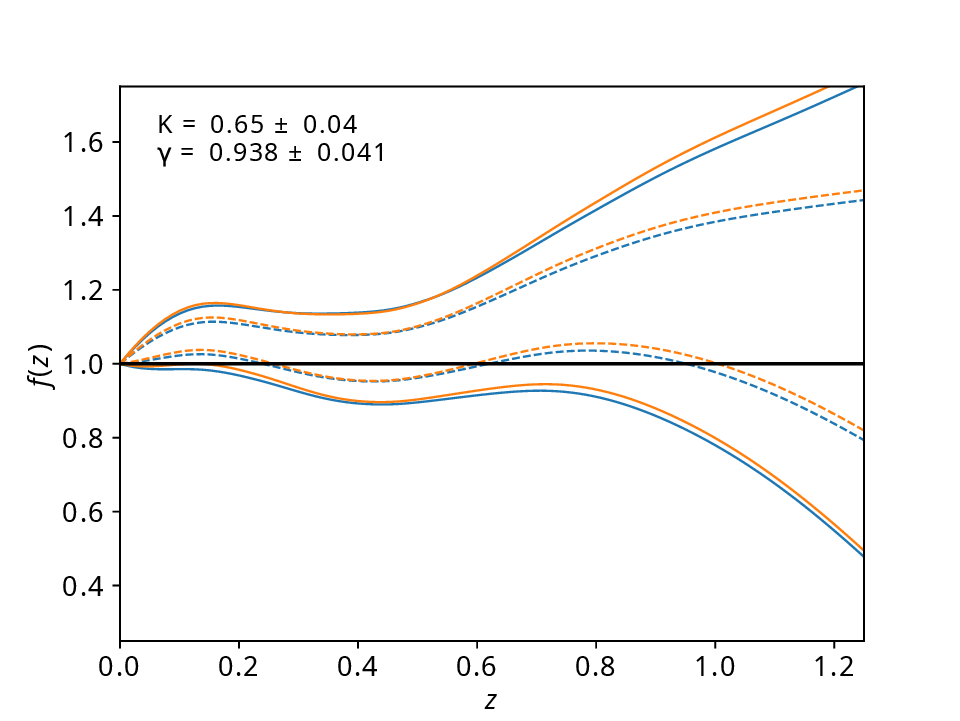}
        \label{fig:comb1}
    \end{subfigure}
    \hfill
    \begin{subfigure}[b]{0.45\textwidth}
        \centering
        \includegraphics[width=\textwidth]{ 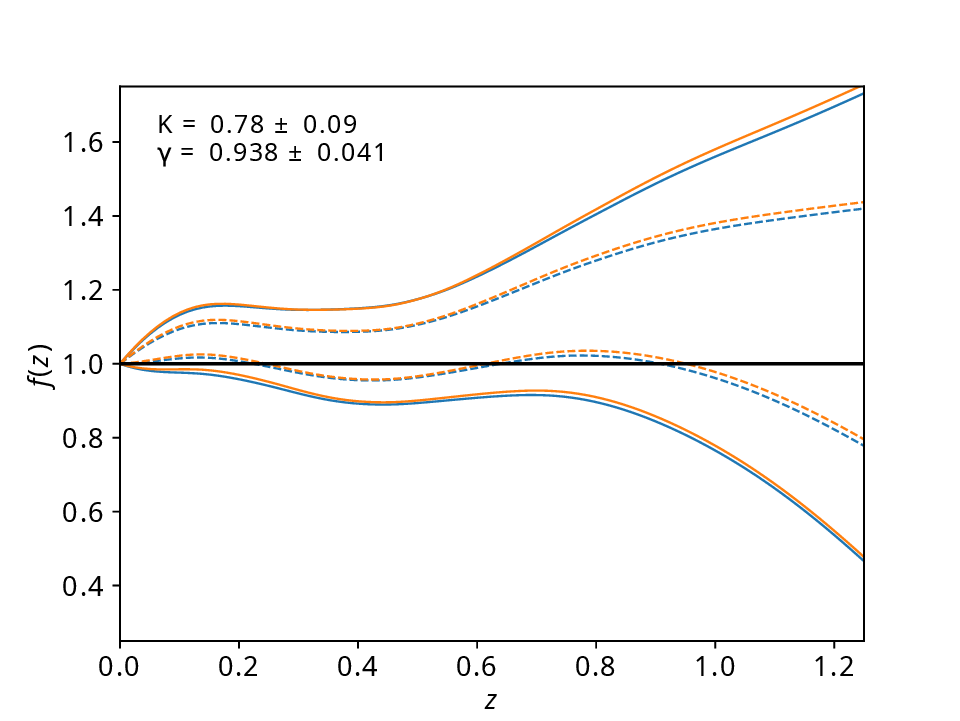}
        \label{fig:comb2}
    \end{subfigure}

    \vspace{0.5cm}

    \begin{subfigure}[b]{0.45\textwidth}
        \centering
        \includegraphics[width=\textwidth]{ 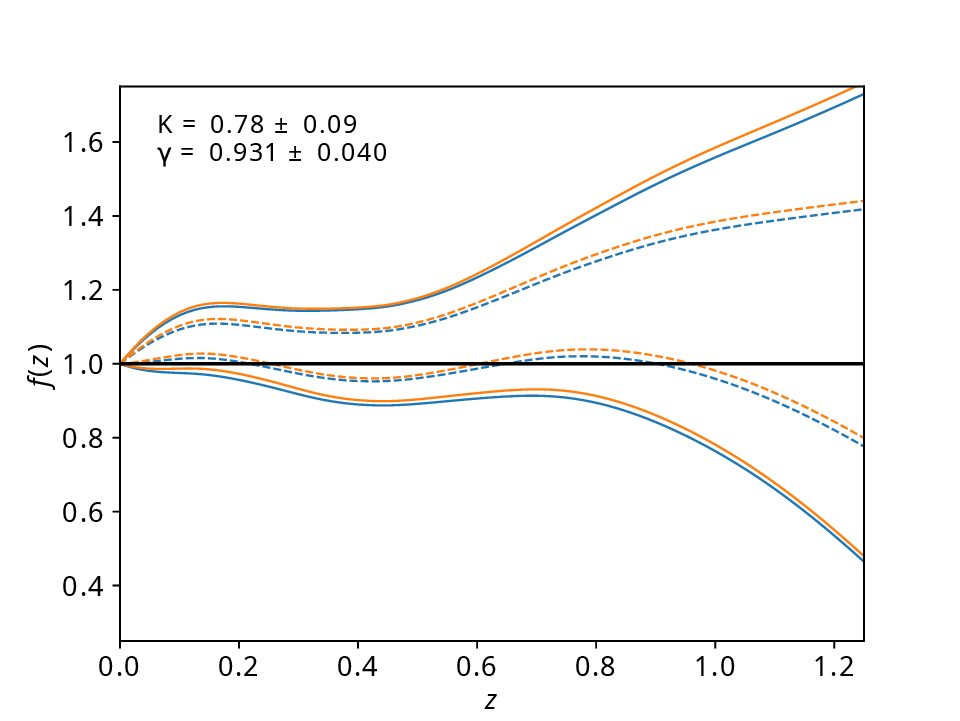}
        \label{fig:comb3}
    \end{subfigure}
    \hfill
    \begin{subfigure}[b]{0.45\textwidth}
        \centering
        \includegraphics[width=\textwidth]{ 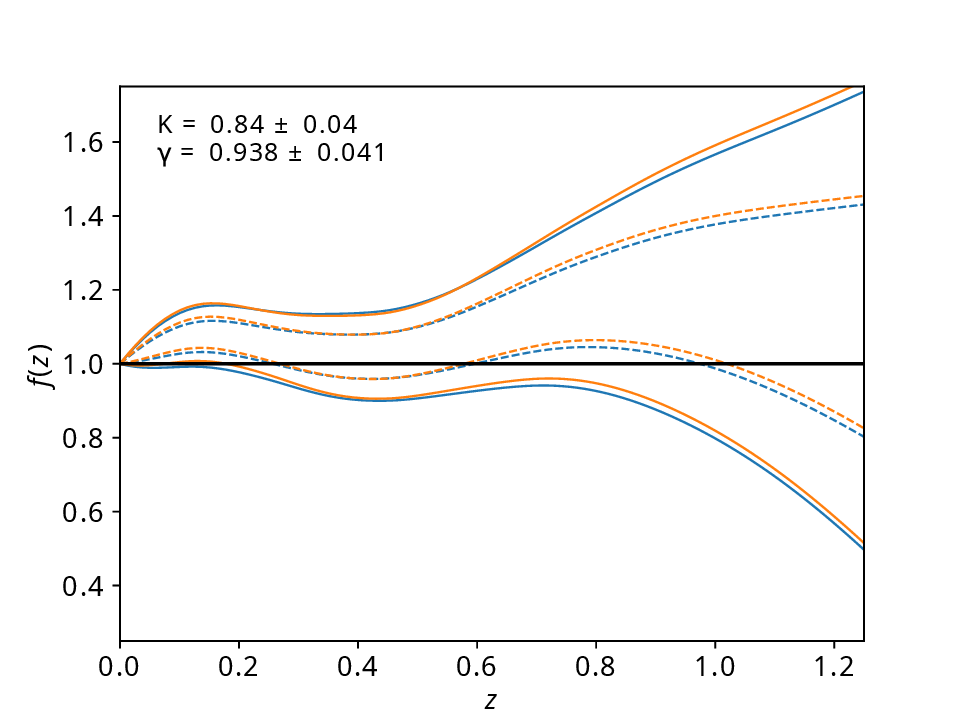}
        \label{fig:comb4}
    \end{subfigure}
    
    \caption{Results of $f(z)$, for each combination of the parameters $\gamma$ and $K$, by using the $D_A$ reconstructed from BAO data,  within the $68\%$ c.l. (dashed curves) and $95\%$ c.l. (solid curves). The orange curves represent the reconstruction considering $h^{\text{SH0ES}}$, while the blue curves represent the cases for $h^{\text{Planck}}$. For reference, the black line represents $f(z) = 1$, the scenario with no deviation.}
    \label{fig:total2}
\end{figure*}

\begin{figure*}
    \centering

    \begin{subfigure}[b]{0.45\textwidth}
        \centering
        \includegraphics[width=\textwidth]{ 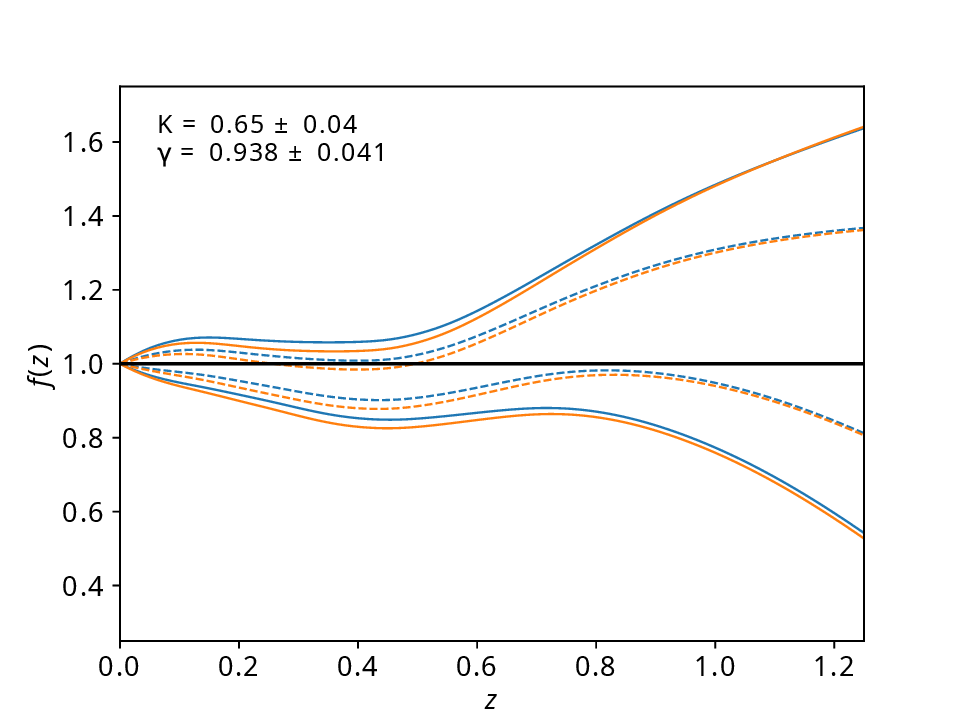}
        \label{fig:1}
    \end{subfigure}
    \hfill
    \begin{subfigure}[b]{0.45\textwidth}
        \centering
        \includegraphics[width=\textwidth]{ 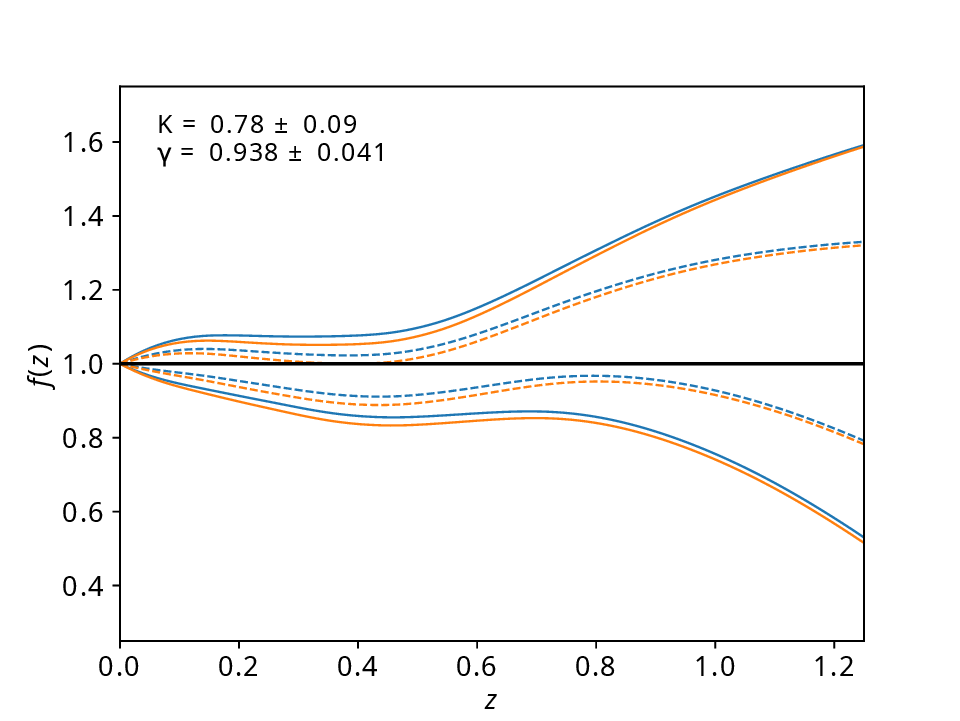}
        \label{fig:2}
    \end{subfigure}

    \vspace{0.5cm}

    \begin{subfigure}[b]{0.45\textwidth}
        \centering
        \includegraphics[width=\textwidth]{ 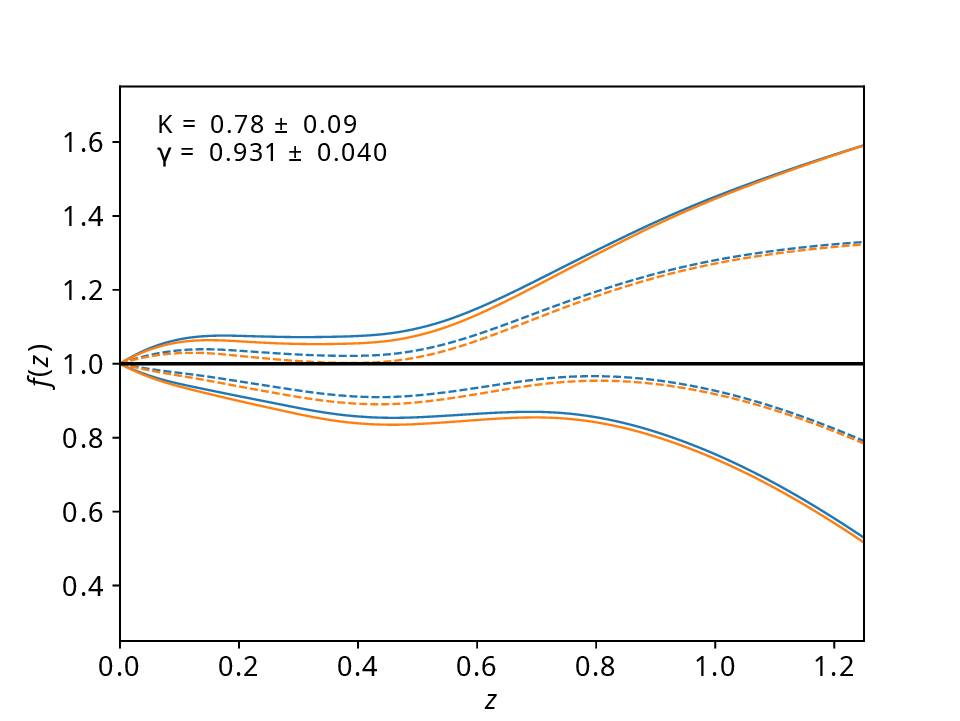}
        \label{fig:3}
    \end{subfigure}
    \hfill
    \begin{subfigure}[b]{0.45\textwidth}
        \centering
        \includegraphics[width=\textwidth]{ 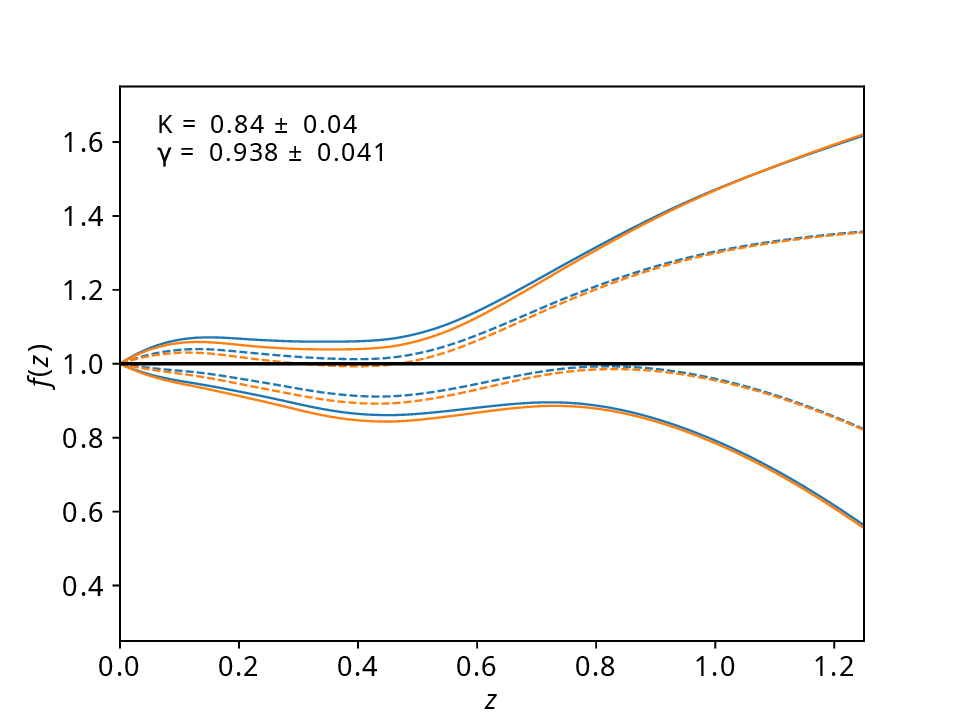}
        \label{fig:4}
    \end{subfigure}
    \caption{ {The $f(z)$ results for each combination of the parameters $\gamma$ and $K$, using the $D_L$ reconstructed from SNe IA data,  within the $68\%$ c.l. (dashed curves) and $95\%$ c.l. (solid curves). The orange curves represent the reconstruction considering $h^{\text{SH0ES}}$, while the blue curves represent the cases for $h^{\text{Planck}}$. For reference, the black line represents $f(z) = 1$, the scenario with no deviation.}}
    \label{fig:panel}
\end{figure*}

Regarding $f(z)$ behaviour, the results are presented in Fig. \ref{fig:total2} and Fig. \ref{fig:panel} by using BAO data and type Ia supernovae measurements, respectively. It can be observed, for both cases,  that the reconstruction aligns with $f(z) = 1$, representing the case of no interaction within $2\sigma$ c.l. across the entire range of data points. {  No notable difference emerged between using $h^{\text{Planck}}$ and $h^{\text{SH0ES}}$ across various combinations of $K$ and $\gamma$, or when employing BAO versus supernova data in the analysis.}  Possibly, this occurs because galaxy cluster data still has relatively high uncertainties (around $15\%$). However, the uncertainties in the $f(z)$ reconstruction do not allow us to exclude the scenario $f(z) \neq 1$ with a high statistical confidence level.  {For illustrative purposes, we present the values of $f(z)$ at redshifts $z = 0.3$, $0.6$, and $1.1$. These values represent points where the reconstructed function exhibits different behaviors. The results for each analysis are shown in Tables \ref{TabResul1} and \ref{TabResul2}.}

  

\begin{figure*}
    \centering
    \begin{subfigure}[b]{0.45\textwidth}
        \centering
        \includegraphics[width=\textwidth]{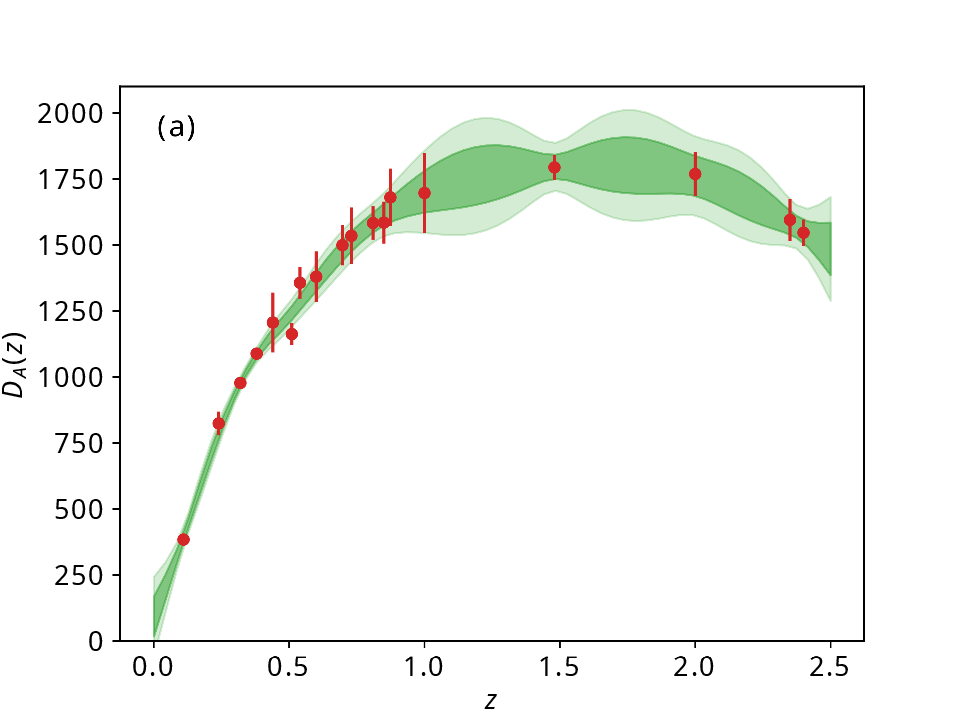}
        \label{fig:comb1}
    \end{subfigure}
    \hfill
    \begin{subfigure}[b]{0.45\textwidth}
        \centering
        \includegraphics[width=\textwidth]{ 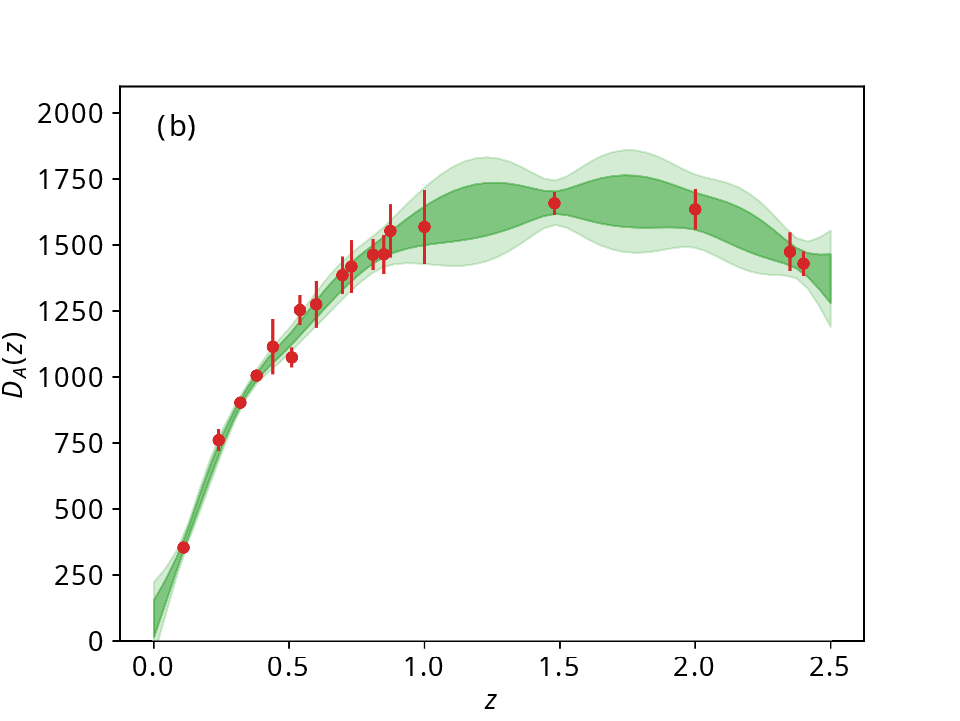}
        \label{fig:comb2}
    \end{subfigure}

    \vspace{0.5cm}

    \begin{subfigure}[b]{0.45\textwidth}
        \centering
        \includegraphics[width=\textwidth]{ 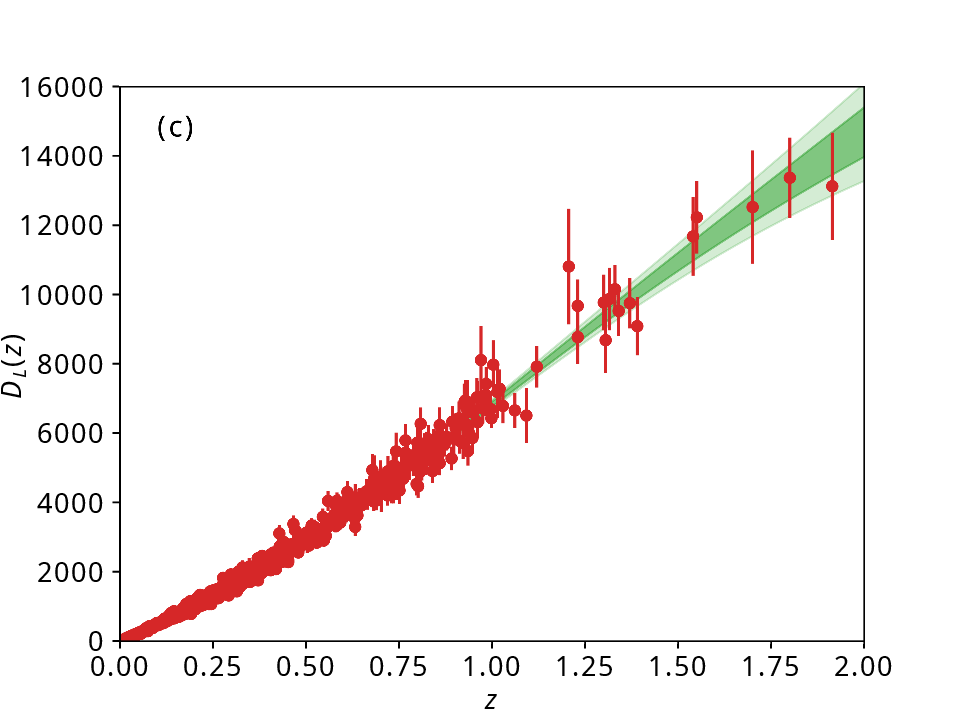}
        \label{fig:comb3}
    \end{subfigure}
    \hfill
    \begin{subfigure}[b]{0.45\textwidth}
        \centering
        \includegraphics[width=\textwidth]{ 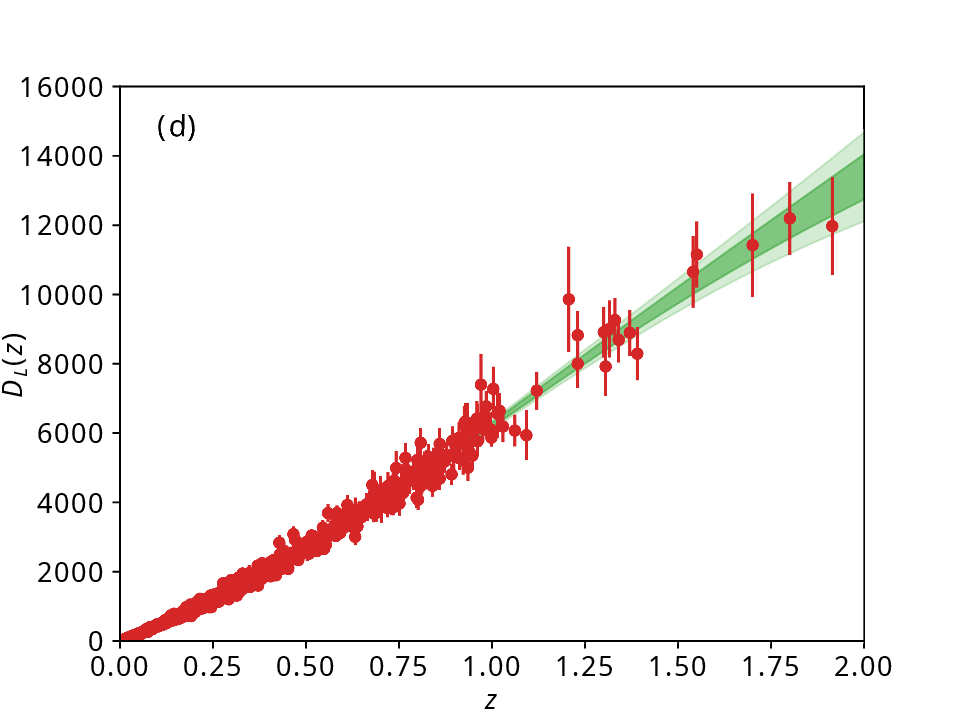}
        \label{fig:comb4}
    \end{subfigure}
    
    \caption{Panel with the four reconstructions for the distances. Figures (a) and (b) show the functions for the angular diameter distance for the cases $h^{\text{Planck}}$ and $h^{\text{SH0ES}}$. Figures (c) and (d) show the functions for the luminosity distance for $h^{\text{Planck}}$ and $h^{\text{SH0ES}}$, respectively. The red points represent the data used for Gaussian process regression, while the dark green and light green areas represent the $68\%$ and $95\%$ confidence levels, respectively.}
    \label{FigDistancesRecons}
\end{figure*}

\begin{table}
\centering
\begin{tabular}{|@{}c@{}|c|c|c|}
\hline
Combination & $f(z=0.3)$ & $f(z=0.6)$ & $f(z=1.1)$ \\
\hline
$K^{\text{CMB}}$ $\gamma^{\text{The300}}$$h^{\text{Planck}}$ & $1.04 \pm 0.05$ & $1.08 \pm 0.08$ & $1.19 \pm 0.25$ \\
\hline
$K^{\text{CLASH}}$ $\gamma^{\text{The300}}$$h^{\text{Planck}}$ & $1.04 \pm 0.06$ & $1.08 \pm 0.08$ & $1.16 \pm 0.25$ \\
\hline
$K^{\text{CLASH}}$ $\gamma^{\text{FABLE}}$$h^{\text{Planck}}$ & $1.04 \pm 0.06$ & $1.08 \pm 0.08$ & $1.17 \pm 0.25$ \\
\hline
$K^{\text{CCCP}}$ $\gamma^{\text{The300}}$$h^{\text{Planck}}$ & $1.04 \pm 0.05$ & $1.09 \pm 0.07$ & $1.19 \pm 0.24$ \\

\hline
$K^{\text{CMB}}$ $\gamma^{\text{The300}}$$h^{\text{SH0ES}}$ & $1.03 \pm 0.05$ & $1.07 \pm 0.08$ & $1.16 \pm 0.25$ \\
\hline
$K^{\text{CLASH}}$ $\gamma^{\text{The300}}$$h^{\text{SH0ES}}$ & $1.03 \pm 0.06$ & $1.07 \pm 0.08$ & $1.15 \pm 0.25$ \\
\hline
$K^{\text{CLASH}}$ $\gamma^{\text{FABLE}}$$h^{\text{SH0ES}}$ & $1.03 \pm 0.06$ & $1.07 \pm 0.08$ & $1.14 \pm 0.25$ \\
\hline
$K^{\text{CCCP}}$ $\gamma^{\text{The300}}$$h^{\text{SH0ES}}$ & $1.03 \pm 0.05$ & $1.08 \pm 0.08$ & $1.16 \pm 0.24$ \\
\hline
\end{tabular}
\caption{Estimated values for $f(z)$ at $z = 0.3$, $0.6$, and $1.1$ obtained from the reconstruction made with BAO data at a $68\%$ confidence level.}
\label{TabResul1}
\end{table}

  

\begin{table}
\centering
\begin{tabular}{|@{}c@{}|c|c|c|}
\hline
Combination & $f(z = 0.3)$ & $f(z=0.6)$ & $f(z=1.1)$ \\
\hline
$K^{\text{cmb}}$ $\gamma^{\text{The300}}$$h^{\text{Planck}}$ & $0.95 \pm 0.05$ & $0.99 \pm 0.07$ & $1.12 \pm 0.22$ \\
\hline
$K^{\text{CLASH}}$ $\gamma^{\text{The300}}$$h^{\text{Planck}}$ & $0.94 \pm 0.05$ & $0.99 \pm 0.08$ & $1.12 \pm 0.25$\\
\hline
$K^{\text{CLASH}}$ $\gamma^{\text{FABLE}}$$h^{\text{Planck}}$ & $0.96 \pm 0.05$ & $0.99 \pm 0.07$ & $1.09 \pm 0.22$ \\
\hline
$K^{\text{CCCP}}$ $\gamma^{\text{The300}}$$h^{\text{Planck}}$ & $0.96 \pm 0.04$ & $0.99 \pm 0.07$ & $1.12 \pm 0.21$ \\
\hline
$K^{\text{CMB}}$ $\gamma^{\text{The300}}$$h^{\text{SH0ES}}$ & $0.97 \pm 0.05$ & $1.01 \pm 0.07$ & $1.12 \pm 0.22$ \\
\hline
$K^{\text{CLASH}}$ $\gamma^{\text{The300}}$$h^{\text{SH0ES}}$ & $0.96 \pm 0.05$ & $1.01 \pm 0.08$ & $1.13 \pm 0.25$ \\
\hline
$K^{\text{CLASH}}$ $\gamma^{\text{FABLE}}$$h^{\text{SH0ES}}$ & $0.98 \pm 0.05$ & $1.01 \pm 0.07$ & $1.09 \pm 0.21$ \\
\hline
$K^{\text{CCCP}}$ $\gamma^{\text{The300}}$$h^{\text{SH0ES}}$ & $0.98 \pm 0.04$ & $1.01 \pm 0.07$ & $1.12 \pm 0.21$ \\

\hline

\end{tabular}
\caption{Estimated values for $f(z)$ using SNe Ia at $z = 0.3$, $0.6$, and $1.1$ at a $68\%$ confidence level.}
\label{TabResul2}
\end{table}


	\section{conclusions}\label{sec7}

Several interaction models have been proposed to capture possible interactions among components of the dark sector of the universe. The interaction typically follows different parametrizations for interaction parameter ($\epsilon (z)$). In this paper, however, considering an approach not dependent on the cosmological model, using only observable data and without considering any parametrization of interaction parameter $\epsilon (z)$, we implement the Gaussian process regression to analyze a possible interaction between dark matter and energy. Specifically, the interaction follows a possible deviation from the standard dark matter density evolution law. This possibility is investigated using galaxy cluster data, BAO measurements, and type Ia supernovae. A generic function representing this hypothesis was non-parametrically reconstructed using Gaussian process regression. For this purpose, we utilized a sample of 103 gas mass fraction data from galaxy clusters (see Fig. \ref{datafgas}), 18 Baryon Acoustic Oscillation data points, and 1048 type Ia supernovae (see Fig. \ref{FigDistancesRecons}). In BAO measurements and SNe Ia observations we considered the two central $h$ values presented in literature:  $h^{\text{SH0ES}} = 0.7324 \pm 0.0174$ (value obtained in the SH0ES project \cite{riess20162}) and $h^{\text{Planck}} = 0.6736 \pm 0.0054$ (from the Planck Collaboration \cite{aghanim2020planck}).
Naturally, the cosmic distance duality relation was used in our analysis with galaxy clusters and SNe Ia to obtain $D_A$ for each galaxy cluster. The BAO and SNe Ia data covered a more extensive redshift range than the $f_{gas}$ data, enabling robust estimates.

 We tested a set of parameters ($h$, $K$, and $\gamma$) which appears in the equation for the gas mass fraction (see Sec. \ref{sec3}). In all scenarios, $f(z)$ exhibited behavior consistent within $2\sigma$ with the standard model (see Figs. \ref{fig:total2} and \ref{fig:panel}). {  No significant difference was observed between using $h^{\text{Planck}}$ and $h^{\text{SH0ES}}$ across various combinations of $K$ and $\gamma$, or when using BAO data compared to supernova data in the analysis (see Tables II and III). }However, the results did not significantly rule out a potential departure from the dark matter density standard evolution. 

In the coming years, there will be a significant increase in the quantity and quality of observations facilitated by instruments such as the Russian-German collaboration telescope eROSITA. This advancement will allow a more robust statistical analysis. As a result, it will be possible to obtain more conclusive results regarding the behavior of dark matter using the methodology employed in this article. Naturally, better galaxy cluster data may show different behaviors for $f(z)$ when different $H_0$ values are used.  Finally, as mentioned earlier, we do not assume a specific parametrization for $\epsilon(z)$. However, an approach considering some phenomenological parametrizations is under investigation and will be reported in a forthcoming communication.

\vspace{0.5cm}

\section*{Acknowledgements}
 Z. C. Santana Junior acknowledges the Coordenação de Aperfeiçoamento de Pessoal de Nível Superior (CAPES) for financial support. R. F. L. Holanda (309132/2020-7) and R. Silva acknowledge the financial support from the Conselho Nacional de Desenvolvimento Científico e Tecnológico (CNPq).

\section{Data Availability Statement}
Data will be made available on reasonable request.

\newpage

\bibliographystyle{ieeetr}
\bibliography{sample.bib}

\begin{thebibliography}{10}

\bibitem{riess1998observational}
A.~G. Riess {\em et~al.} {\em AJ}, vol.~116, p.~1009, 1998.

\bibitem{perlmutter1999measurements}
S.~Perlmutter {\em et~al.} {\em ApJ}, vol.~517, p.~565, 1999.

\bibitem{weinberg2013observational}
D.~H. Weinberg, M.~J. Mortonson, D.~J. Eisenstein, C.~Hirata, A.~G. Riess, and E.~Rozo {\em Phys. report}, vol.~530, pp.~87--255, 2013.

\bibitem{ozer1986possible}
M.~{\"O}zer and M.~Taha {\em Phys. Lett. B}, vol.~171, pp.~363--365, 1986.

\bibitem{sahni2005dark}
V.~Sahni in {\em The Physics of the Early Universe}, pp.~141--179, 2004.

\bibitem{caldera2009growth}
G.~Caldera-Cabral, R.~Maartens, and B.~M. Schaefer {\em JCAP}, vol.~2009, p.~027, 2009.

\bibitem{ade2016planck}
P.~A. Ade {\em et~al.} {\em Astron. Astrophys.}, vol.~594, p.~A13, 2016.

\bibitem{kamionkowski2023hubble}
M.~Kamionkowski and A.~G. Riess {\em Annu. Rev. Nucl.}, vol.~73, pp.~153--180, 2023.

\bibitem{di2021realm}
E.~Di~Valentino, O.~Mena, S.~Pan, L.~Visinelli, W.~Yang, A.~Melchiorri, D.~F. Mota, A.~G. Riess, and J.~Silk {\em Class Quantum Gravity}, vol.~38, p.~153001, 2021.

\bibitem{weinberg1989cosmological}
S.~Weinberg {\em RMP}, vol.~61, p.~1, 1989.

\bibitem{padmanabhan2003cosmological}
T.~Padmanabhan {\em Phys. Rep.}, vol.~380, pp.~235--320, 2003.

\bibitem{lombriser2023cosmology}
L.~Lombriser {\em Class Quantum Gravity}, 2023.

\bibitem{zlatev1999quintessence}
I.~Zlatev, L.~Wang, and P.~J. Steinhardt {\em Phys. Rev. Lett.}, vol.~82, p.~896, 1999.

\bibitem{perivolaropoulos2022challenges}
L.~Perivolaropoulos and F.~Skara {\em New Astron. Rev.}, vol.~95, p.~101659, 2022.

\bibitem{elcio2022cosmology}
A.~Elcio {\em et~al.} {\em JHEAP}, vol.~34, pp.~49--211, 2022.

\bibitem{joyce2015beyond}
A.~Joyce, B.~Jain, J.~Khoury, and M.~Trodden {\em Phys. Rep.}, vol.~568, pp.~1--98, 2015.

\bibitem{wang2016dark}
B.~Wang, E.~Abdalla, F.~Atrio-Barandela, and D.~Pavon {\em Rep. Prog. Phys.}, vol.~79, p.~096901, 2016.

\bibitem{amendola2000coupled}
L.~Amendola {\em Phys. Rev. D}, vol.~62, p.~043511, 2000.

\bibitem{carroll2001cosmological}
S.~M. Carroll {\em Living Rev. Relativ.}, vol.~4, pp.~1--56, 2001.

\bibitem{nelson1982scaling}
B.~L. Nelson and P.~Panangaden {\em Phys. Rev. D}, vol.~25, p.~1019, 1982.

\bibitem{elizalde1994renormalization}
E.~Elizalde and S.~D. Odintsov {\em Phys. Lett. B}, vol.~321, pp.~199--204, 1994.

\bibitem{elizalde1995gut}
E.~Elizalde, C.~O. Lousto, S.~D. Odintsov, and A.~Romeo {\em Phys. Rev. D}, vol.~52, p.~2202, 1995.

\bibitem{bytsenko1994effective}
A.~Bytsenko, S.~Odintsov, and S.~Zerbini {\em Phys. Lett. B}, vol.~336, pp.~355--361, 1994.

\bibitem{wagoner1970scalar}
R.~V. Wagoner {\em Phys. Rev. D}, vol.~1, p.~3209, 1970.

\bibitem{linde1974lee}
A.~D. Linde {\em JETP Lett.}, vol.~19, p.~183, 1974.

\bibitem{polyakov1982phase}
A.~Polyakov, ``P,'' {\em Sov. Phys.}, vol.~25, p.~187, 1982.

\bibitem{cohen1999effective}
A.~G. Cohen, D.~B. Kaplan, and A.~E. Nelson {\em Phys. Rev. Lett.}, vol.~82, p.~4971, 1999.

\bibitem{shapiro2002scaling}
I.~L. Shapiro and J.~Sola {\em J. of High Energy Phys.}, vol.~2002, p.~006, 2002.

\bibitem{shapiro2000scaling}
I.~L. Shapiro and J.~Sola {\em Phys. Lett. B}, vol.~475, pp.~236--246, 2000.

\bibitem{shapiro2003variable}
I.~L. Shapiro, J.~Sola, C.~Espana-Bonet, and P.~Ruiz-Lapuente {\em Phys. Lett. B}, vol.~574, pp.~149--155, 2003.

\bibitem{espana2004testing}
C.~Espana-Bonet, P.~Ruiz-Lapuente, I.~L. Shapiro, and J.~Sola {\em J. Cosmol. Astropart. Phys.}, vol.~2004, p.~006, 2004.

\bibitem{weinberg1993vacuum}
E.~J. Weinberg {\em Phys. Rev. D}, vol.~47, p.~4614, 1993.

\bibitem{sola2011cosmologies}
J.~Sola in {\em JPCS}, vol.~283, p.~012033, 2011.

\bibitem{alcaniz2005interpreting}
J.~S. Alcaniz and J.~A. S.~d. Lima {\em Phys. Rev. D}, vol.~72, p.~063516, 2005.

\bibitem{jesus2022can}
J.~Jesus, A.~Escobal, D.~Benndorf, and S.~Pereira {\em Eur. Phys. J. C}, vol.~82, p.~273, 2022.

\bibitem{pereira2009can}
S.~Pereira and J.~Jesus {\em Phys. Rev. D}, vol.~79, p.~043517, 2009.

\bibitem{bolotin2015cosmological}
Y.~L. Bolotin, A.~Kostenko, O.~A. Lemets, and D.~A. Yerokhin {\em Int. J. Mod. Phys. D}, vol.~24, p.~1530007, 2015.

\bibitem{amendola2007consequences}
L.~Amendola, G.~C. Campos, and R.~Rosenfeld {\em Phys. Rev. D}, vol.~75, p.~083506, 2007.

\bibitem{tamanini2015phenomenological}
N.~Tamanini {\em Phys. Rev. D}, vol.~92, p.~043524, 2015.

\bibitem{nunes2022new}
R.~C. Nunes, S.~Vagnozzi, S.~Kumar, E.~Di~Valentino, and O.~Mena {\em Phys. Rev. D}, vol.~105, p.~123506, 2022.

\bibitem{yang2019dark}
W.~Yang, O.~Mena, S.~Pan, and E.~Di~Valentino {\em Phys. Rev. D}, vol.~100, p.~083509, 2019.

\bibitem{lucca2020shedding}
M.~Lucca and D.~C. Hooper {\em Phys. Rev. D}, vol.~102, p.~123502, 2020.

\bibitem{liu2022dark}
Y.~Liu, S.~Liao, X.~Liu, J.~Zhang, R.~An, and Z.~Fan {\em MNRAS}, vol.~511, pp.~3076--3088, 2022.

\bibitem{wang2004can}
P.~Wang and X.-H. Meng {\em Class Quantum Gravity}, vol.~22, p.~283, 2004.

\bibitem{bora2022test}
K.~Bora, R.~Holanda, S.~Desai, and S.~Pereira {\em Eur. Phys. J. C}, vol.~82, p.~17, 2022.

\bibitem{holanda2019estimate}
R.~Holanda, R.~Gon{\c{c}}alves, J.~Gonzalez, and J.~Alcaniz {\em JCAP}, vol.~2019, p.~032, 2019.

\bibitem{bora2021probing}
K.~Bora, R.~Holanda, and S.~Desai {\em Eur. Phys. J. C}, vol.~81, pp.~1--7, 2021.

\bibitem{von2019cosmological}
R.~von Marttens, L.~Casarini, D.~Mota, and W.~Zimdahl {\em Phys. Dark Universe}, vol.~23, p.~100248, 2019.

\bibitem{cid2019bayesian}
A.~Cid, B.~Santos, C.~Pigozzo, T.~Ferreira, and J.~Alcaniz {\em JCAP}, vol.~2019, p.~030, 2019.

\bibitem{von2020unphysical}
R.~von Marttens, H.~Borges, S.~Carneiro, J.~Alcaniz, and W.~Zimdahl {\em Eur. Phys. J. C}, vol.~80, p.~1110, 2020.

\bibitem{costa2010cosmological}
F.~Costa and J.~Alcaniz {\em Phys. Rev. D}, vol.~81, p.~043506, 2010.

\bibitem{costa2010coupled}
F.~Costa {\em Phys. Rev. D}, vol.~82, p.~103527, 2010.

\bibitem{da2020thermodynamic}
W.~da~Silva, J.~Gonzalez, R.~Silva, and J.~Alcaniz {\em Eur. Phys. J. Plus}, vol.~135, pp.~1--11, 2020.

\bibitem{Gonzalez2018}
J.~E. Gonzalez, H.~H.~B. Silva, R.~Silva, and J.~S. Alcaniz {\em Eur. Phys. J. C}, vol.~78, p.~730, 2018.

\bibitem{allen2011cosmological}
S.~W. Allen, A.~E. Evrard, and A.~B. Mantz {\em ARAA}, vol.~49, pp.~409--470, 2011.

\bibitem{qiu2023cosmology}
L.~Qiu, N.~R. Napolitano, S.~Borgani, F.~Zhong, X.~Li, M.~Radovich, W.~Lin, K.~Dolag, C.~Tortora, Y.~Wang, {\em et~al.} {\em arXiv preprint arXiv:2304.09142}, 2023.

\bibitem{chaubal2022improving}
P.~Chaubal {\em et~al.} {\em ApJ}, vol.~931, p.~139, 2022.

\bibitem{mantz2022cosmological}
A.~B. Mantz {\em et~al.} {\em MNRAS}, vol.~510, pp.~131--145, 2022.

\bibitem{corasaniti2021cosmological}
P.-S. Corasaniti, M.~Sereno, and S.~Ettori, ``C,'' {\em ApJ}, vol.~911, p.~82, 2021.

\bibitem{wu2021cosmology}
H.-Y. Wu, D.~H. Weinberg, A.~N. Salcedo, and B.~D. Wibking {\em ApJ}, vol.~910, p.~28, 2021.

\bibitem{holanda2020low}
R.~Holanda, G.~Pordeus-da Silva, and S.~Pereira {\em JCAP}, vol.~2020, p.~053, 2020.

\bibitem{lesci2022amico}
G.~Lesci {\em et~al.} {\em Astron. Astrophys}, vol.~665, p.~A100, 2022.

\bibitem{sasaki1996new}
S.~Sasaki {\em PASJ}, vol.~48, pp.~L119--L122, 1996.

\bibitem{allen2008improved}
S.~Allen, D.~Rapetti, R.~Schmidt, H.~Ebeling, R.~Morris, and A.~Fabian {\em MNRAS}, vol.~383, pp.~879--896, 2008.

\bibitem{mantz2014cosmology}
A.~B. Mantz, S.~W. Allen, R.~G. Morris, D.~A. Rapetti, D.~E. Applegate, P.~L. Kelly, A.~von~der Linden, and R.~W. Schmidt {\em MNRAS}, vol.~440, pp.~2077--2098, 2014.

\bibitem{battaglia2013cluster}
N.~Battaglia, J.~Bond, C.~Pfrommer, and J.~Sievers {\em ApJ}, vol.~777, p.~123, 2013.

\bibitem{applegate2016cosmology}
D.~Applegate, A.~Mantz, S.~Allen, A.~v. der Linden, R.~G. Morris, S.~Hilbert, P.~L. Kelly, D.~Burke, H.~Ebeling, D.~Rapetti, {\em et~al.} {\em MNRAS}, vol.~457, pp.~1522--1534, 2016.

\bibitem{holanda2017cosmological}
R.~Holanda, V.~Busti, J.~Gonzalez, F.~Andrade-Santos, and J.~Alcaniz {\em JCAP}, vol.~2017, p.~016, 2017.

\bibitem{eckert2019non}
D.~Eckert {\em et~al.} {\em A\&A}, vol.~621, p.~A40, 2019.

\bibitem{ettori2010mass}
S.~Ettori, F.~Gastaldello, A.~Leccardi, S.~Molendi, M.~Rossetti, D.~Buote, and M.~Meneghetti {\em A\&A}, vol.~524, p.~A68, 2010.

\bibitem{ghirardini2017evolution}
V.~Ghirardini, S.~Ettori, S.~Amodeo, R.~Capasso, and M.~Sereno {\em A\&A}, vol.~604, p.~A100, 2017.

\bibitem{cooke2018one}
R.~J. Cooke, M.~Pettini, and C.~C. Steidel {\em ApJ}, vol.~855, p.~102, 2018.

\bibitem{riess20162}
A.~G. Riess, L.~M. Macri, S.~L. Hoffmann, D.~Scolnic, S.~Casertano, A.~V. Filippenko, B.~E. Tucker, M.~J. Reid, D.~O. Jones, J.~M. Silverman, {\em et~al.} {\em ApJ}, vol.~826, p.~56, 2016.

\bibitem{aghanim2020planck}
N.~Aghanim, Y.~Akrami, M.~Ashdown, J.~Aumont, C.~Baccigalupi, M.~Ballardini, A.~Banday, R.~Barreiro, N.~Bartolo, S.~Basak, {\em et~al.} {\em A\&A}, vol.~641, p.~A6, 2020.

\bibitem{staicova2022constraining}
D.~Staicova and D.~Benisty {\em A\&A}, vol.~668, p.~A135, 2022.

\bibitem{arendse2020cosmic}
N.~Arendse, R.~J. Wojtak, A.~Agnello, G.~C.-F. Chen, C.~D. Fassnacht, D.~Sluse, S.~Hilbert, M.~Millon, V.~Bonvin, K.~C. Wong, {\em et~al.} {\em A\&A}, vol.~639, p.~A57, 2020.

\bibitem{scolnic2018complete}
D.~M. Scolnic, D.~Jones, A.~Rest, Y.~Pan, R.~Chornock, R.~Foley, M.~Huber, R.~Kessler, G.~Narayan, A.~Riess, {\em et~al.} {\em ApJ}, vol.~859, p.~101, 2018.

\bibitem{camarena2021use}
D.~Camarena and V.~Marra {\em MNRAS}, vol.~504, pp.~5164--5171, 2021.

\bibitem{foreman2013emcee}
D.~Foreman-Mackey, D.~W. Hogg, D.~Lang, and J.~Goodman {\em PASP}, vol.~125, p.~306, 2013.

\bibitem{gelman1992inference}
A.~Gelman and D.~B. Rubin {\em Stat Sci}, vol.~7, pp.~457--472, 1992.

\bibitem{williams2006gaussian}
C.~K. Williams and C.~E. Rasmussen, vol.~2.
\newblock MIT press Cambridge, MA, 2006.

\bibitem{wang2023intuitive}
J.~Wang {\em Computing in Science \& Engineering}, 2023.

\bibitem{schulz2018tutorial}
E.~Schulz, M.~Speekenbrink, and A.~Krause {\em Journal of mathematical psychology}, vol.~85, pp.~1--16, 2018.

\bibitem{sereno2015comparing}
M.~Sereno and S.~Ettori {\em MNRAS}, vol.~450, pp.~3633--3648, 2015.

\bibitem{herbonnet2020cccp}
R.~Herbonnet, C.~Sif{\'o}n, H.~Hoekstra, Y.~Bah{\'e}, R.~F. van Der~Burg, J.-B. Melin, A.~von Der~Linden, D.~Sand, S.~Kay, and D.~Barnes {\em MNRAS}, vol.~497, pp.~4684--4703, 2020.

\bibitem{hoekstra2015canadian}
H.~Hoekstra, R.~Herbonnet, A.~Muzzin, A.~Babul, A.~Mahdavi, M.~Viola, and M.~Cacciato {\em MNRAS}, vol.~449, pp.~685--714, 2015.

\bibitem{salvati2018constraints}
L.~Salvati, M.~Douspis, and N.~Aghanim {\em A\&A}, vol.~614, p.~A13, 2018.

\bibitem{cui2018three}
W.~Cui, A.~Knebe, G.~Yepes, F.~Pearce, C.~Power, R.~Dave, A.~Arth, S.~Borgani, K.~Dolag, P.~Elahi, {\em et~al.} {\em MNRAS}, vol.~480, pp.~2898--2915, 2018.

\bibitem{henden2020baryon}
N.~A. Henden, E.~Puchwein, and D.~Sijacki {\em MNRAS}, vol.~498, pp.~2114--2137, 2020.

\end{thebibliography}

\end{document}